\documentclass[draft]{agujournal2019}
\usepackage{subcaption}

\usepackage[finalnew]{trackchanges}

\addeditor{a}
\addeditor{r1}
\addeditor{r2}

\usepackage{soul}

\usepackage{changepage}

\draftfalse

\journalname{JGR: Space Physics}

\begin{document}

\title{Coordinate Systems and Transforms in Space Physics: Terms, Definitions, Implementations, and Recommendations for Reproducibility}

\authors{R.S. Weigel\affil{1}\, A.Y. Shih\affil{2}\, R. Ringuette\affil{3}\, I. Christopher\affil{4}\, S.M. Petrinec\affil{5}\, S. Turner\affil{6}\, R.M. Candey\affil{7}\, G.K. Stephens\affil{6}\, and B. Cecconi\affil{8}}

\affiliation{1}{George Mason University, Space Weather Lab, Fairfax, VA, USA}
\affiliation{2}{NASA/GSFC, Greenbelt, MD, USA}
\affiliation{3}{University of Maryland, Baltimore County, Baltimore, Maryland, USA}
\affiliation{4}{University of Iowa, Iowa City, IA, USA}
\affiliation{5}{Lockheed Martin Advanced Technology Center, Palo Alto, CA, USA}
\affiliation{6}{JHU Applied Physics Laboratory, Laurel, MD, USA}
\affiliation{7}{NASA/GSFC, Greenbelt, MD, USA}
\affiliation{8}{Observatoire de Paris, Université de Recherche Paris Sciences et Lettres, Paris, France}

\noindent

\correspondingauthor{R.S. Weigel}{rweigel@gmu.edu}

\begin{keypoints}
\item Reproducing coordinate system transforms is difficult due to inconsistencies in definitions and implementations in software.
\item Spacecraft positions from different providers can differ significantly, as do $Z$-axis orientations from surveyed software libraries.
\item We present recommendations to improve reproducibility and reduce effort duplication by missions.
\end{keypoints}

\begin{abstract}
In space physics, acronyms for coordinate systems (e.g., \texttt{GEI}, \texttt{GSM}) are commonly used; however, differences in their definitions and implementations can prevent reproducibility. In this work, we compare definitions in online resources, software packages, and frequently cited journal articles and show that implementation differences can lead to transformations between same-named coordinate systems and position values from different data providers to differ significantly. Based on these comparisons and results, and to enable reproducibility, we recommend that (a) a standard for acronyms and definitions for coordinate systems is developed, similar to equivalents in astronomy or earth sciences; (b) a standards body develops a citable database of reference data needed for these transforms. For software that computes coordinate transforms, we also recommend that their developers provide explicit comparisons of their implementations with the results of (b) and documentation on implementation choices. Additionally, we provide recommendations for scientists and metadata developers to ensure that sufficient information is provided to enable reproducibility. Finally, we document that spacecraft positions from data providers can differ both because of differences in how they implemented transforms and because of differences in the original source of the position data, and provide recommendations to improve the documentation of spacecraft positional datasets.

\end{abstract}

\section*{Plain Language Summary}

To facilitate reproducibility in analyses that use measurements transformed into different coordinate systems, standards for the names and definitions of coordinate systems are needed. We motivate this by showing that different data providers and software can produce transforms between same-named coordinate systems that significantly differ. A commonly transformed measurement is spacecraft position. We show that positions from data providers can differ both due to differences in how they implemented transforms and because of differences in the original source of the position data. Based on these observations, we provide a set of recommendations to improve reproducibility for researchers using transformed data and spacecraft position values.

\section{Introduction}

In space physics, coordinate systems are usually referenced by an acronym. The coordinate system's definition is typically assumed or given by citing a reference that describes it. This work stems from a project to develop a standard for terms and descriptions for coordinate system acronyms such as \texttt{GEI}, \texttt{GSE}, \texttt{GSM}, and their commonly used variations, e.g., \texttt{GEI\_J2000}, \texttt{GEI\_MOD}; see \ref{sect:glossary} for definitions and \citeA{Russell1971}, \citeA{Hapgood1992}, \citeA{Franz2002}, and \citeA{Laundal2016} for additional details. The objective is that, given a statement such as ``the vector measurements in \texttt{ABC} were transformed into \texttt{XYZ},'' a scientist with measurements in \texttt{ABC} would be able to reproduce the transformation, with the only uncertainty being due to rounding errors associated with floating-point precision.

Currently, this is not possible due to the lack of standards. At present, the best that can be achieved is a citation to the definition in one of the papers above, but this is not sufficient for reproducibility -- in this work we show that how a given coordinate frame definition can be implemented in different ways, leading to different data providers giving different locations for a spacecraft and different software libraries giving different answers for the transformation of a vector from one coordinate system to another.

As documented in this work, comparing data transformed with different software is complicated by the fact that transform implementations are not unique, introducing unknown uncertainty. In the early era of space physics, uncertainty due to implementation differences was much smaller than measurement uncertainty. Here, we argue that, in some applications, these differences matter and that implementation uncertainty poses an unnecessary impediment to reproducibility and data intercomparison. In addition, as a matter of scientific practice, terms and definitions should be precise.

Our background research revealed additional issues that must be addressed to meet the objective of reproducibility in the sense defined above. First, implementing a given coordinate system given a general description requires making implementation choices. Therefore, terms and definitions for reference implementations are also needed. In section~\ref{sect:terminology}, we note that in astronomy and the Earth sciences, this has been recognized and addressed, and we describe their terminology and conventions.

Second, the standard for terms and descriptions must address their existing ambiguity. Examples of ambiguity are given for the \texttt{GEI} coordinate system in section~\ref{sect:definitions}. The general conclusion is that in space physics, there is no agreed-upon standard for coordinate-system acronyms, and descriptions generally lack key implementation details.

To quantify the impact of ambiguity in definitions and implementations, we give examples that demonstrate the reported positions for spacecraft in the same-named reference systems can significantly differ depending on the data source in section~\ref{sect:comparisons_ephemeris}, and that the source of these differences can be due to differences in the definition of a coordinate system. An additional issue arose from this analysis. We also show that different data providers report different spacecraft positions, so even when both providers use the same coordinate system definition, results may differ depending on the source of the position data. Examples of this are given in section~\ref{sect:comparisons_ephemeris}.

In section~\ref{sect:comparisons_software}, we demonstrate that software that transforms a vector between two coordinate systems exhibits smaller, but possibly significant, differences.

Having summarized in section~\ref{sect:comparisons} the problems arising from ambiguity in definitions and implementations in practice, we next consider how spacecraft missions compute transforms. To meet the objective of reproducibility, we must understand how calculations are performed and account for this in our recommendations. In section~\ref{sect:missions}, we describe how missions develop data products in different coordinate systems, which may explain the differences observed in section~\ref{sect:comparisons_ephemeris}.

In section~\ref{sect:conclusions}, we provide a set of recommendations that address definition issues described in section~\ref{sect:definitions}, the results in section~\ref{sect:comparisons}, and the description in section~\ref{sect:missions}. These recommendations address ambiguity in terminology and definitions, the creation and testing of software for transforms, and how missions compute transformed data products.

\section{Coordinate Systems, Reference Systems, and Reference Frames}
\label{sect:terminology}

Thus far, we have used the term ``coordinate system" consistent with common usage in space physics as described in section~\ref{sect:coordsystems}. In section~\ref{sect:refsystems}, we note that ``ideal reference system", ``reference system", or ``reference frame" is more consistent with literature outside of space physics and will use these terms in the remainder of this article.

\subsection{Coordinate Systems}
\label{sect:coordsystems}

In geometry, introductory physics, and mathematics textbooks, coordinate values in a ``coordinate system" uniquely identify spatial positions relative to an {\it arbitrary} set of three orthogonal vectors and an origin. Common coordinate systems are Cartesian, cylindrical, and spherical. In the space physics literature, a coordinate system is generally understood to be a {\it specific} set of three orthogonal vectors and an origin, with coordinate values expressed in a coordinate system such as Cartesian, cylindrical, or spherical.

This usage, in which ``coordinate system" has two meanings, dates back to at least early documentation of the NASA shuttle program \cite{Davis1974} and has been consistently used in frequently cited literature related to space physics reference systems (\citeA{Russell1971}; \citeA{Hapgood1992}; \citeA{Hapgood1995}; \citeA{Laundal2016}). (However, an earlier work involving analysis of data from the OGO (Orbiting Geophysical Observatory; \citeA{NASA1970}) satellite used only ``system" for the central noun, e.g., \texttt{GCI} system, instead of \texttt{GCI} coordinate system.)

\subsection{Reference Systems and Frames}
\label{sect:refsystems}

In astronomy, the terms ``reference system" and ``reference frame" are used; from \citeA{USNOICRS}:

``A reference system is the complete specification of how a celestial coordinate system is to be formed. It defines the origin and fundamental planes (or axes) of the coordinate system. It also specifies all of the constants, models, and algorithms used to transform between observable quantities and reference data that conform to the system. A reference frame consists of a set of identifiable fiducial points on the sky (specific astronomical objects), along with their coordinates, that serves as the practical realization of a reference system."

The terms ``reference system" and ``reference frame" are also used in the same sense in terrestrial geodesy \cite{Seitz2014}.

A fundamental reference system in astronomy is the International Celestial Reference System (ICRS) \cite{Petit2010}. No unique reference frame is associated with this reference system because creating (or, equivalently, ``realizing" or ``implementing") one requires measurements for computing reference system model parameters. The International Celestial Reference Frame (ICRF) is the general term for realizations of the ICRS that are adopted by a standards body and updated as new measurements and model versions become available. There are three versions of the ICRF \cite{Charlot2020}. 

In the Earth sciences, an example of a fundamental reference system is the World Geodetic System 1984 (WGS~84), used to represent locations on Earth in terms of latitude, longitude, and altitude (\citeA{NGA2014}). A standard for expressing machine-readable reference frame implementation details that facilitates reproducibility is also available \cite{OGC2023}.

In the Earth sciences, \citeA{Kovalevsky1981} and \citeA{Mueller1985} use the term ``ideal reference system'' to refer to reference frames with definitions that are incomplete in the sense that only fundamental planes, axes, and an origin are specified: ``The term `ideal' indicates the conceptual definition only and that no means are proposed to actually construct the system". Note that the term ``ideal'' in this quote should be interpreted not as meaning ``preferred" but rather means ideal in the sense of ``idealized model", or a model that has practical or important attributes omitted to simplify its description.



Thus, what space physicists call a ``coordinate system'' (e.g., \texttt{GEI}, \texttt{GSM}) is analogous to an ``ideal reference system'' because it only has a general definition for \change{their}{its} axes orientation and origin. 
There are no standards for the constants, models, or algorithms required to transform between observable quantities, which are necessary to define a reference system, nor for the model parameters needed to define a reference frame. 
For example, the transformation of the \texttt{GEI} reference frame to \texttt{GEO} requires a rotation by the Greenwich Mean Sidereal Time (GMST) ---
\citeA{Franz2002} cites an equation from \citeA{Meeus1998}, which also appears in \citeA{USNO1992}; \citeA{Hapgood1992} uses a reduced-accuracy and reduced-order GMST equation that is based on an equation in \citeA{USNO1989}; both use J2000.0 as the epoch. \citeA{Russell1971} gives a Fortran program for GMST that uses December 31, 1899, at 12:00 UT as the epoch and was provided through private communication. The equation used appears to be a reduced-accuracy and reduced-order version of that in \citeA{Newcomb1898} \add[r2]{(p.~ 9)}; see also \citeA{HMNAO1961} \add[r2]{(pp.~74-75)} and \citeA{Hapgood1995}. Thus, \citeA{Russell1971}, \citeA{Hapgood1992}, and \citeA{Franz2002} each define unique \texttt{GEI} to \texttt{GEO} reference frame transforms.

\section{Examples of Definition Ambiguity}
\label{sect:definitions}

In the previous section, examples were given of general issues with the terminology used to describe reference systems and frames. For specific reference systems or frames in space physics, there are additional ambiguities.

For example, consider \texttt{GEI}, which is based on the astronomical reference system that specifies locations using right ascension and declination. The \texttt{GEI} ideal reference frame has $\mathbf{Z}$ aligned with Earth's rotation axis with positive northward, $\mathbf{X}$ as the intersection of Earth's equatorial plane with the plane of Earth's orbit around the Sun (the ecliptic plane), with positive in the direction from the Earth to the Sun at the time of the vernal equinox, and $\mathbf{Y}=\mathbf{Z}\times\mathbf{X}$. The line of intersection of Earth's equatorial plane with the ecliptic, or the time that it is computed, is sometimes referred to as ``the equinox," and whether a line or time is meant must be determined from context.

First, there is no consistency in the expansion of the acronym \texttt{GEI}. \citeA{Russell1971} and \citeA{Hapgood1992} associate \texttt{GEI} with ``Geocentric Equatorial Inertial System". \citeA{Franz2002} and \citeA{SunPy} associate \texttt{GEI} with ``Geocentric Earth Equatorial". Another ambiguity is in the expansion of \texttt{GCI}; \citeA{SSCWeb} notes ``Geocentric Inertial (\texttt{GCI}) and Earth-Centered Inertial (\texttt{ECI}) are the same as \texttt{GEI}.", which implicitly defines \texttt{GCI} as ``Geocentric Inertial'', in contrast to \citeA{Russell1971} and
\citeA{NASA1970}, which use ``Geocentric Celestial Inertial''.

When reporting positions for \texttt{GEI}, the origin is taken as the center of mass of Earth. To establish this as a reference system as defined in section~\ref{sect:refsystems}, Earth's rotation axis, center, and the ecliptic plane must be specified, along with the models used to compute them. To define a \texttt{GEI} reference frame, the model parameters must also be specified.

Common variations of \texttt{GEI} depend on whether precession and nutation of Earth's rotation axis and precession of the ecliptic plane are accounted for (\citeA{Davis1974}; \citeA{Hapgood1995}; \citeA{Franz2002}). The term ``mean epoch--of--date" (or mean--of--date) is used when only precession is accounted for; that is, the nutation variation is estimated and removed. If nutation variation is not removed, the term ``true epoch--of--date" or ``true--of--date" is used \cite{Hapgood1995}. If the $\mathbf{X}$ and $\mathbf{Z}$ orientations are time-independent, an abbreviation for a reference date and time is specified, e.g., J2000.0 (2000-01-01T12:00 Terrestrial Time), at which they were determined. The reference date and time are referred to as an ``epoch" in the sense of a reference time instant rather than a period of time.

Two basic categories of \texttt{GEI}-related reference frames and systems are commonly used: inertial and non-inertial. The list of acronyms later in this section is associated with an inertial reference system: an idealized system that is not rotating with respect to the distant stars, i.e., stars at an effectively infinite distance from its origin, and with an origin that translates with a constant velocity \cite{NAIFOverview2023}. Note that if the origin is specified as Earth's center (usually its center of mass rather than its centroid is implied but not stated; see also \citeA{Dong2003}), \texttt{GEI} is non-inertial by definition. However, in the bulleted list summary of \texttt{GEI}--related reference systems and frames below, we have ignored this technicality.

The objective of the following summary is to demonstrate the diversity in the terms and definitions used for \texttt{GEI}--related reference systems and to motivate the need for the standard recommended in section~\ref{sect:standard-for}. Similar diversity also exists in terminology and definitions in other commonly used reference systems in space physics, providing further motivation.

\begin{itemize}
    \parskip 0.1in 

    \item \texttt{J2000} -- Used by Satellite Situation Center System and Services (2025c) to refer to a frame with its origin at Earth's center of mass using the reference epoch J2000.0 (2000 January 1 noon Terrestrial Time).

     \citeA{SSCWeb} provides the definition:

     ``Geocentric Equatorial Inertial for epoch J2000.0 (\texttt{GEI2000}), also known as Mean Equator and Mean Equinox of J2000.0 (Julian date 2451545.0 TT (Terrestrial Time), or 2000 January 1 noon TT, or 2000 January 1 11:59:27.816 TAI or 2000 January 1 11:58:55.816 UTC.) This system has X-axis aligned with the mean equinox for epoch \texttt{J2000}; Z-axis is parallel to the rotation axis of the Earth, and Y completes the right-handed orthogonal set.''

    For the SPICE (Spacecraft, Planet, Instrument, C-matrix, Events; \citeA{Acton1997}, \citeA{NAIFGeneral2025}) library, \citeA{NAIFFrames2025} notes that \texttt{J2000} is ``generally used in SPICE to refer to the \texttt{ICRF}"; ``The rotational offset between the \texttt{J2000} frame and the \texttt{ICRS} has magnitude of under 0.1 arcseconds [$2.\overline{7}\cdot 10^{-5}$ degrees]."; ``The \texttt{ICRF} frame is defined by the adopted locations of 295 extragalactic radio sources."; ``The \texttt{J2000} (aka \texttt{EME2000}) frame definition is based on the Earth’s equator and equinox, determined from observations of planetary motions, plus other data.''; and ``The realization of \texttt{ICRF} was made to coincide almost exactly with the \texttt{J2000} frame.'' 

    \item \texttt{GEI2000} -- \citeA{SSCWeb} identifies this as equivalent to \texttt{J2000} by a parenthetical statement in the definition of \texttt{J2000}: ``Geocentric Equatorial Inertial for epoch J2000.0 (\texttt{GEI2000})."

    \item \texttt{GeocentricEarthEquatorial} -- Used by SunPy \cite{SunPy} with supporting definition of ``A coordinate or frame in the Geocentric Earth Equatorial (\texttt{GEI}) system."

    \item \texttt{GCRS} --  Used in Astropy \cite{AstroPy2022} with supporting definition ``A coordinate or frame in the Geocentric Celestial Reference System (\texttt{GCRS}). \texttt{GCRS} is distinct from \texttt{ICRS} mainly in that it is relative to the Earth’s center-of-mass rather than the solar system Barycenter. That means this frame includes the effects of aberration (unlike \texttt{ICRS})."

    \item \texttt{GEI}$_\mathrm{J2000}$ -- \citeA{Franz2002} state that \texttt{GEI}$_\mathrm{J2000}$ is realized through the \texttt{ICRF}. This is ambiguous today, because there are now three \texttt{ICRF} versions \cite{Charlot2020}.

    \item \texttt{EME2000} -- Earth Mean Equator and Equinox \texttt{J2000.0}; defined in \citeA{NAIFFrames2025} via ``... with those of the \texttt{J2000} (aka \texttt{EME2000}) reference frame". \citeA{SSCWeb} defines as ``Mean Equator and Mean Equinox of \texttt{J2000.0}" in its definition of \texttt{J2000}.

    \item \texttt{ECI2000} -- Earth Centered Inertial for \texttt{J2000} epoch; used in SpacePy \cite{Niehof2022, SpacePy}.

    \item \texttt{ECI} -- Earth Centered Inertial; used in \citeA{Burch2026} with the note ``\texttt{GEI/J2000} - Earth-Centered Inertial (\texttt{ECI}). To fully specify the system the equinox must be defined. This system uses the mean equinox at the J2000 epoch."

   \item \texttt{GCI} -- Geocentric Celestial Inertial; used in OGO satellite data analysis \cite{NASA1970} and stated in \citeA{Russell1971} as equivalent to \texttt{GEI}.

\end{itemize}

\section{Comparisons}
\label{sect:comparisons}

Space physics researchers and analysts have several options for obtaining data in different coordinate frames or transforming between them.

\begin{enumerate}

    \parskip 0.1in

    \item From datasets provided to a data archive, e.g., NASA's Coordinated Data Analysis Web (CDAWeb) and ESA's Cluster Science Archive (CSA), with vector measurements in different reference systems; in this case, a dataset in a needed reference frame is selected, if available.

    \item Using data services that provide vector measurements in different reference frames. For example, SSCWeb provides positions for spacecraft in various reference frames as a function of time \cite{SSCWeb} and a calculator that takes an input of a geocentric position in a given reference system at a specific time and outputs the position in other reference systems \cite{SSCWebCoordinateCalculator}.

    \item Using software packages that have coordinate transform functions. For example, in Python, Astropy \cite{AstroPy2022}, SunPy \cite{SunPy}, SpacePy \cite{SpacePy}, and PySPEDAS \cite{PySPEDAS}. Other libraries include \texttt{cxform} \cite{cxform} in C, and Geopack-2008 \cite{Tsyganenko2008} and IRBEM \cite{IRBEM2022} in Fortran.

\end{enumerate}

It is sometimes argued that, although reference-frame implementations may yield different results, this uncertainty is small relative to measurement uncertainty. For example, \citeA{Hapgood1995} notes that ``The attitude error that arises from an error in the epoch used to compute \texttt{GEI} is $0.036^\circ$, which is small compared to the angular resolution of most space plasma measurements ...". 

Although this statement was made thirty years ago, based on the results in this section given below, it appears the convention in space physics is that differences at this level are considered negligible. That is, we find angular differences in spacecraft positions from different providers of ${\sim}0.3^\circ$ and coordinate transforms from different software packages of ${\sim} 0.03^\circ$, and this uncertainty is not discussed in the documentation, suggesting it is considered small enough to be ignored.

Although the convention for determining which uncertainties are considered negligible may have been appropriate at one time, we suggest revisiting these assumptions.

First, spacecraft constellations have achieved separations that correspond to an angular separation relative to Earth's center of ${\sim} 0.003^\circ$ --- based on SSCWeb data, MMS-2 and MMS-3 had a separation of 4.5~ km at 2016-09-15T21:47:30Z, corresponding to an angular separation of $0.0034^\circ$. Cluster 3 and Cluster 4 were separated by 2.4~km at 2018-12-02T00:31:30Z, corresponding to an angular separation of $0.0029^\circ$ (see also \citeA{Escoubet2021}).

Second, having different data providers or software libraries compute transforms in different ways can impede validation, for example, 

\begin{itemize}
\item In an attempt to develop a 3D visualization program that returns the region of geospace (e.g., plasma sheet, interplanetary medium, plasmasphere) given an arbitrary position, we implemented the models used by SSCWeb \cite{SSCWeb} and attempted to validate our work by comparing our predicted regions at the positions provided by its web service for spacecraft. Although our predictions were generally consistent, we were unable to reproduce the exact time at which a spacecraft is predicted to cross from one region to another. After a search for errors in our model implementation failed to explain the differences, we realized that the discrepancies were due to the software that we used for reference-frame transformation differing from that used by SSCWeb. Although the uncertainty in the models is much larger than the uncertainty due to the differences in the transform implementations, the transform differences led to difficulty in validating that our implementation of an idealized model matched that of SSCWeb. That is, ideally, one does not want to report that an implementation matches the original implementation except for errors that could be due to differences in how reference frame transforms are implemented.

\item Not all global magnetosphere simulation models are executed in the same reference frame. To compare two models, the solar wind input must be transformed to the simulation reference frame, and simulated variables from both models must be transformed into the same reference frame for comparison. The uncertainty associated with transformations must be known to determine the difference in the model output attributable to the simulation algorithm. 
\end{itemize}

The fact that different software libraries provide different options for reference frames is indirect evidence that differences can matter. For example, SpacePy \cite{SpacePy} provides multiple options for computing the \texttt{MAG} system, which have angular differences of ${\sim}0.01^\circ$, and SunPy \cite{SunPy} has options for how the \texttt{GEI} frame is computed. A general finding is that differences in transform results within a single library, with different options for the transform, are on the order of, or smaller than, differences in transform results between two different software libraries. As a result, improvements in how a transform is computed within a single library are lost when comparing data computed with a different library. 

Even if the uncertainty associated with how a given reference frame was implemented is small relative to that of the measurements, we suggest that it is illogical for two data providers to report the location of a spacecraft that differs by more than what is expected due to numerical precision unless the reason for differences is clearly documented and ideally evident in the name of the dataset, e.g., by including the word ``preliminary". A recommendation for addressing this is given in section~\ref{sect:ephemeris-version}.

\subsection{Spacecraft Positions}
\label{sect:comparisons_ephemeris}

In this section, we give examples showing how position values from data providers can differ. We quantify the differences in terms of three quantities: 

\begin{itemize}
    \item $|\Delta\mathbf{r}|/R_E$, the magnitude of the difference in position vectors relative to an Earth radius, with $R_E\equiv 6378.16$~km, which is the value used by SSCWeb (see discussion in section~\ref{sect:R_E} about the use of $R_E$);
    \item $|\Delta\mathbf{r}|/\overline{r}$ the magnitude of the difference in position vectors relative to the average of the magnitudes of the position vectors; and
    \item  $\Delta \theta$, the angular difference between position.
\end{itemize}

All vectors in these quantities have an origin at the center of Earth. 

\noindent Angular differences can be compared to:
\begin{enumerate}
    \item The average angular change in Earth's dipole $Z$ axis of ${\sim} 0.04^\circ$ per year (from 1970--2025, computed using the libraries used in section~\ref{sect:comparisons_software}).

    \item The precession, also referred to as luni--solar precession, of Earth's rotation axis (also the \texttt{GEI}$_Z$ reference system axis), which drifts by ${\sim} 0.006^\circ$ per year \cite{Hapgood1995}.

    \item The nutation of Earth's rotation axis of ${\sim} 0.0025^\circ$ per year \cite{Hapgood1995}.

    \item The precession of the ecliptic (planetary precession) of $\Delta \theta {\sim} 0.000014^\circ$ per year \cite{Hapgood1995}.

    \item The angular separation of ${\sim}0.003^\circ$ achieved between spacecraft in a multi-spacecraft constellation.
\end{enumerate}

\begin{figure}[htb]
    \vspace{-0.5cm}
    \includegraphics[width=1.1\textwidth]{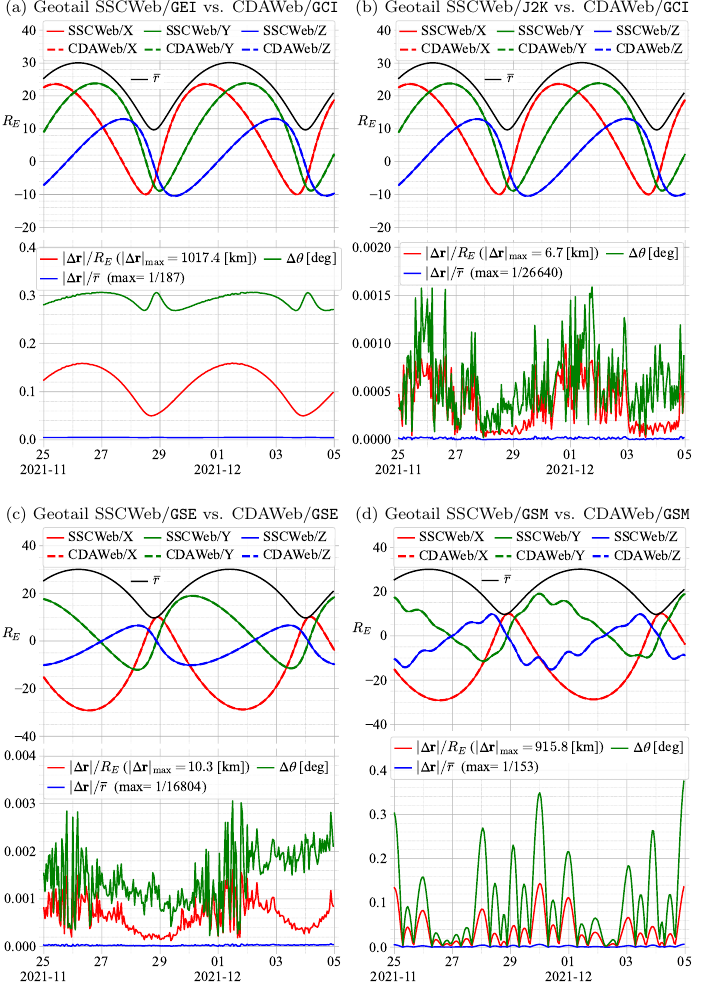}
    \caption{Comparison of position values for the Geotail spacecraft from SSCWeb and CDAWeb in four different reference systems. The top panel in each subplot displays the $X$, $Y$, and $Z$ values from each provider (on this scale, differences are not visible) and the average radial distance, $\overline{r}$, between the two providers. The bottom panel of each subplot shows relative differences in the position vector $\Delta\mathbf{r}$ and the angular difference in the position vectors, $\Delta\theta$.}
    \label{fig:geotail}
\end{figure}

Figure~\ref{fig:geotail} shows the position of the Geotail spacecraft from 2021-11-25 through 2021-12-05 from the SSCWeb web service and the GE\_OR\_DEF dataset in CDF \cite{CDF2025} files from CDAWeb. SSCWeb provides positions for scientific spacecraft in reference systems that include \texttt{GEI}, \texttt{J2K}, \texttt{GSE}, \texttt{GSM}, and \texttt{GM} (frequently referred to as \texttt{MAG}). The GE\_OR\_DEF dataset has position values in \texttt{GCI}, \texttt{GSE}, and \texttt{GSM}.

We used the SSCWeb option to return data as a fraction of $R_E$ with 10 fractional digits. The CDAWeb dataset stores values in km as IEEE-754 64-bit floats, and the metadata indicates a recommended display precision of 10 fractional digits. The cadence of the SSCWeb Geotail data is 12~minutes, while the CDAWeb Geotail dataset is 10~minutes. In the time interval displayed in Figure~\ref{fig:geotail}, values from the datasets are only shown when they have the same timestamps in the time interval shown.

Based on the precision of the data, and if both providers use the same definition of \texttt{GEI}, we expect $|\Delta r|/R_E \simeq 10^{-10}$ given that the SSCWeb values are reported in $R_E$ to $10$ fractional digits and the CDAWeb values are recommended for display with 10 fractional digits. For $\overline{r}=20 R_E$, this corresponds to an uncertainty due to the precision of the data of $\Delta \theta \simeq 20\cdot (180/\pi) \cdot 10^{-10}$ degrees $\simeq 10^{-7}$ degrees.

In Figure~\ref{fig:geotail}(a), and based on the documentation for \citeA{SSCWeb}, which has the statement ``Geocentric Inertial (\texttt{GCI}) and Earth-Centered Inertial (\texttt{ECI}) are the same as \texttt{GEI}.", we expect the SSCWeb/\texttt{GEI} and CDAWeb/\texttt{GCI} positions differences to be near $\simeq 10^{-7}$ degrees, assuming both data sources use the same definition of \texttt{GEI}. The average angular difference between position vectors is $\Delta \theta \sim 0.3^\circ$, and the average error relative to an Earth radius, $|\Delta\mathbf{r}|/R_E\sim 0.16$, or 16\%. The maximum error relative to the average geocentric distance, $|\Delta \mathbf{r}|/\overline{r}$ is $1/187$, or $0.5$\%.

In Figure~\ref{fig:geotail}(b), there is a much closer match between SSCWeb/J2K and CDAWeb/GCI positions, with an average angular difference in the position vector of ${\sim}0.001^\circ$ and an average error relative to an Earth radius of ${\sim}0.05$\%. 

In summary, Figure~\ref{fig:geotail}(a) and (b) show that what is labeled as \texttt{GCI} in the CDAWeb positions is much closer to SSCWeb/\texttt{J2K} positions than SSCWeb/\texttt{GEI}, which is unexpected given the SSCWeb documentation that notes \texttt{GCI} and \texttt{GEI} are equivalent. Although the CDAWeb/\texttt{GCI} positions are a better match to the SSCWeb/\texttt{J2K} positions, the differences are not explained by the numerical precision of the reported values. They may be due to whether or how precession and nutation in items 2--4 listed above were accounted for (see \citeA{Hapgood1995} for a related discussion).

In Figure~\ref{fig:geotail}(c), a comparison is made for the \texttt{GSE} reference system. To transform a vector from \texttt{GEI} to \texttt{GSE}, two rotations are required: a rotation around \texttt{GEI}$_X$ to align Earth's equator with the ecliptic plane and a rotation around the \texttt{GEI}$_Z$ axis (as it is after the first rotation) to orient \texttt{GEI}$_X$ to point to center of the Sun. These two angles are the obliquity of the ecliptic and the ecliptic longitude of the Sun, respectively (\citeA{Hapgood1992}).

In Figure~\ref{fig:geotail}(d), a comparison is made for the \texttt{GSM} reference system. A vector in the \texttt{GSE} reference system is transformed to \texttt{GSM} by an angle that depends on the orientation of the North geomagnetic pole (\citeA{Russell1971}; \citeA{Hapgood1992}). The primary uncertainty is how the orientation of the North geomagnetic pole, which has an average yearly change of ${\sim}0.04^\circ$, is computed. If both SSCWeb and CDAWeb computed the \texttt{GSM} positions based on a transform from \texttt{GSE}, we could conclude that the observed $\Delta \theta$ are due to the use of different locations of the North geomagnetic pole. However, the maximum $\Delta \theta \simeq 0.37^\circ$ is much larger than the average yearly angular change in Earth's dipole $Z$ axis of ${\sim}0.04^\circ$.

\begin{figure}[h]
    \vspace{-0.5cm}
    \includegraphics[width=1.1\textwidth]{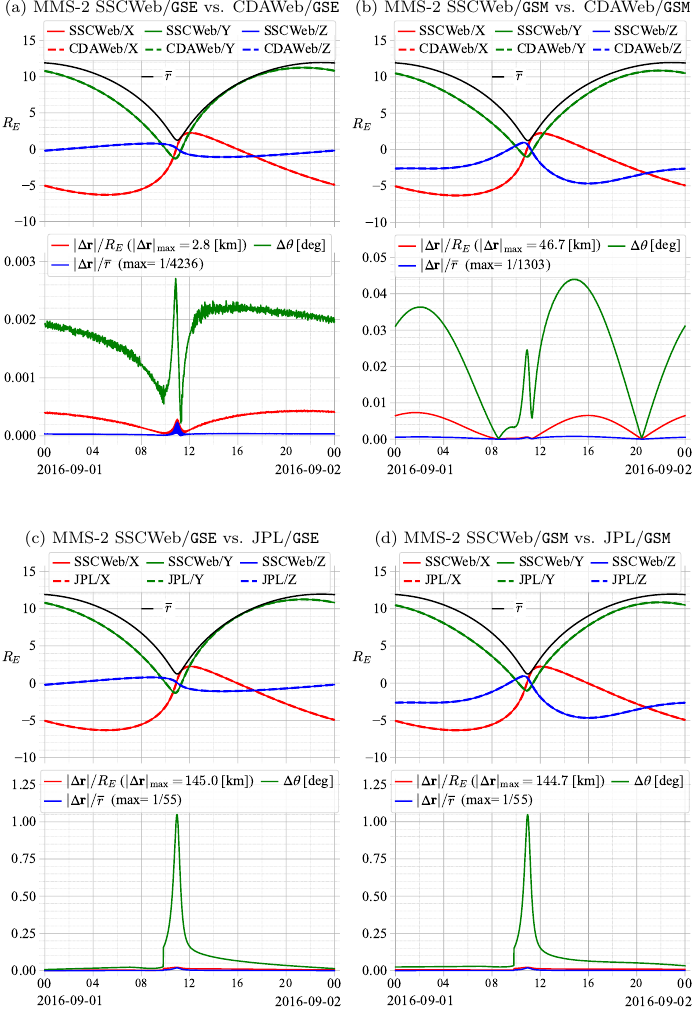}
    \caption{Comparison of position values for the MMS-2 spacecraft from (a)--(b) SSCWeb and CDAWeb; (c)--(d) SSCWeb and JPL Horizons in two different reference systems.}
    \label{fig:mms-2}
\end{figure}

In Figure~\ref{fig:mms-2}, a comparison is made between data from SSCWeb, CDAWeb, and the JPL Horizons web service \cite{JPLHorizons}, which provides positions for some of the spacecraft available from SSCWeb. The CDAWeb dataset is MMS2\_EPD-EIS\_SRVY\_L2\linebreak\_ELECTRONENERGY. The JPL Horizons web service provides positions in the \texttt{ICRF} reference frame (the version of \texttt{ICRF} is not documented), and SunPy is used to transform it to other reference frames.

Figure~\ref{fig:mms-2}a shows that the $\Delta \mathbf{r}$ between \texttt{GSE} locations from CDAWeb and SSCWeb match to within ${\sim}3$~km, which is much larger than expected by numerical precision but smaller than the closest separation distance of MMS spacecraft of ${\sim}7$~km \cite{NASA2016}.

In Figure~\ref{fig:mms-2}c, much larger differences are shown to exist between SSCWeb and JPL than between SSCWeb and CDAWeb, with the maximum angular difference being up to ${\sim}1^\circ$ instead of $0.003^\circ$ found in the comparison of SSCWeb and CDAWeb in Figure~\ref{fig:mms-2}a. A similar conclusion is made by a comparison of Figure~\ref{fig:mms-2}b and Figure~\ref{fig:mms-2}d. A possible explanation for the sharp increase in $\Delta \theta$ near 11:00 is a thruster firing near perigee on this day -- the raw JPL Horizons web service output notes that ``The spacecraft may be maneuvered frequently. Therefore, the JSpOC TLE-based [Joint Space Operations Center Two--Line Element] trajectory provided here may at times depart from the actual trajectory. This can happen because TLEs do not model thruster firings; the TLE trajectory solutions must be reinitialized after each event.''

\clearpage

\clearpage

\clearpage

\subsection{Software}
\label{sect:comparisons_software}

In this section, we compare reference frame calculations using the following software packages. 

\begin{itemize}
    \item Geopack-2008 double precision (\citeA{Tsyganenko2008}; labeled as \texttt{geopack\_08\_dp} in the figures in this section), a Fortran library that was upgraded from earlier versions to support calculations using 64-bit floating-point values.
    \item PySPEDAS 2.1.1, which uses transformation code derived from SPEDAS, which is written in IDL \cite{Angelopoulos2024IDL}; SPEDAS transformations are based on the the ROCOTLIB Fortran \cite{ROCOTLIB} library, which was used for data from the Cluster spacecraft mission \cite{ClusterTools1993}.
    \item SpacePy 0.6.0, which has an option to use the IRBEM Fortran library (\citeA{IRBEM2022}; labeled as \texttt{spacepy-irbem}) or an alternative and native Python implementation (\texttt{spacepy}) of the transforms.
    \item SpiceyPy 8.1.0 \cite{SpiceyPy} is a Python wrapper for the SPICE toolkit. Two versions of SPICE kernel files were used, indicated by \texttt{spicepy1} and \texttt{spicepy2}. \texttt{spicepy1} was used for reference frame transforms for the Van Allen Probes mission. \texttt{spicepy2} had an update to the \texttt{MAG} frame, which used a more recent version of the IGRF model \cite{Alken2021} to determine the Earth's magnetic dipole orientation. (The IGRF model is maintained by the IAGA \cite{IAGA2026} V-MOD \cite{IAGA_V-MOD2026} Working Group.)
    \item SunPy version 7.1.2, which extends the Astropy coordinates framework to include space physics reference frames.
\end{itemize}

\begin{figure}[htb]
    \vspace{-0.65cm}
    \includegraphics[width=\textwidth]{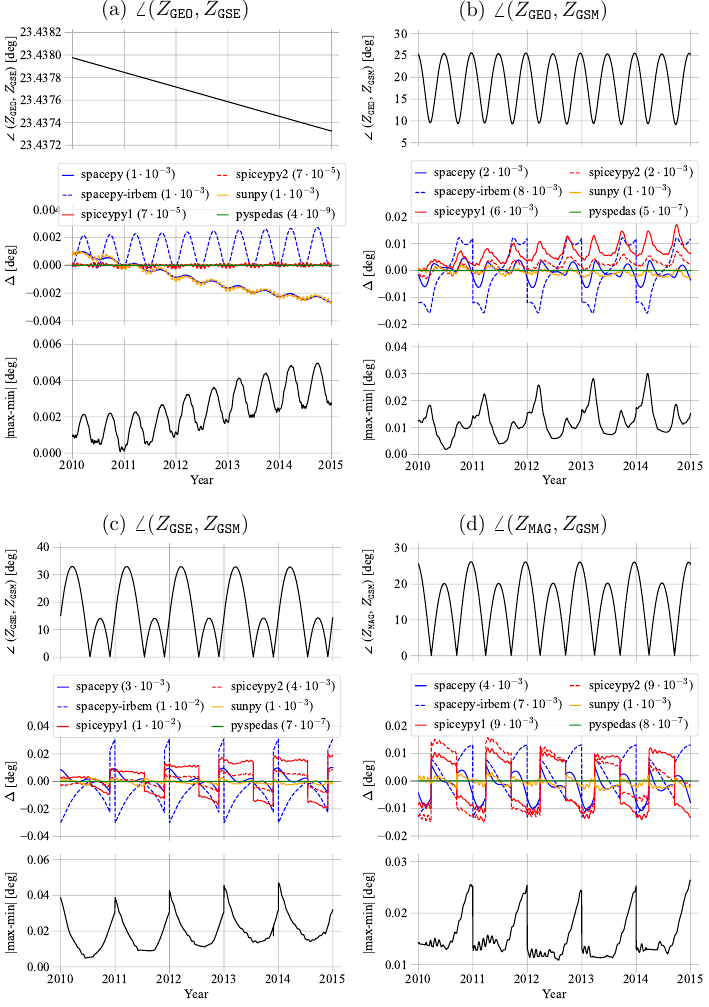}
    \caption{Angles between $Z$ axes in select coordinate frames as computed by different software packages. In the top panels of each subplot, the notation $\angle (Z_\mathtt{A}, Z_\mathtt{B})$ means the angle between the $Z$ axis of coordinate frame $\mathtt{A}$ and the $Z$ axis of coordinate frame $\mathtt{B}$ computed using \texttt{geopack\_08\_dp}. The middle panel of each subplot shows the difference, $\Delta$, between $\angle (Z_\mathtt{A}, Z_\mathtt{B})$ computed using \texttt{geopack\_08\_dp} and that computed using the library indicated in the legend; the maximum of the absolute value of each line is shown in parentheses. The bottom panels show the maximum absolute differences in the middle panel.}
    \label{fig:angles}
\end{figure}

In Figure~\ref{fig:angles}(a)-(d), the top rows show the angles between the $Z$ axis of four coordinate frame pairs as computed by \texttt{geopack\_08\_dp}. The angle between the $Z$ axis in frame \texttt{A} and the $Z$ axis in frame \texttt{B}, $\angle (Z_\mathtt{A}, Z_\mathtt{B})$, was computed by transforming the $Z$ axis in frame \texttt{A} into frame \texttt{B}, and the angle was computed between this transformed axis and the $Z$ axis in frame \texttt{B}. 

The middle rows in Figure~\ref{fig:angles}(a)-(d) show the difference, $\Delta$, in the angles between the $Z$ axes for each library with respect to that for \texttt{geopack\_08\_dp}. The choice of library used as the baseline for this comparison is arbitrary, and it should not be interpreted as implying that this library is preferred; in general, we are interested in the maximum differences, $|\mbox{max}-\mbox{min}|$, among the libraries, which are shown in the bottom rows of Figure~\ref{fig:angles}(a)-(d). The maximum absolute values shown in the bottom rows of Figure~\ref{fig:angles}(a)-(d) range from $4\cdot 10^{-9}$ degrees for \texttt{pyspedas} in Figure~\ref{fig:angles}(a) to $1\cdot 10^{-2}$ degrees for \texttt{spicepy1} in Figure~\ref{fig:angles}(c); the values for each library are given in the legend in the middle row of each subplot. From inspection of the source code for \texttt{geopack\_08\_dp} and \texttt{pyspedas}, we find that they appear to be independent implementations of the algorithms in Russell (1971), which is consistent with the closeness of their transform results shown in the middle rows of Figure~\ref{fig:angles}~(a)-(d).

The bottom rows in Figure~\ref{fig:angles}(a)-(d) shows that across all libraries, the maximum absolute value of lines in the middle rows are ${\sim}0.01^\circ$ for \texttt{GEO}/\texttt{GSE}, ${\sim} 0.03^\circ$ for \texttt{GEO}/\texttt{GSM}, ${\sim}0.04^\circ$ for \texttt{GSE}/\texttt{GSM}, and ${\sim}0.03^\circ$ for \texttt{GSE}/\texttt{GSM}, all of which are comparable to the average angular change in Earth's dipole $Z$ axis of ${\sim}0.04^\circ$ per year.

We have also performed coordinate transforms using the \citeA{SSCWebCoordinateCalculator} coordinate calculator web service. However, this service outputs values only to two decimal places, leading to angular differences of $0.3^\circ$. These values are not plotted so that the other library differences are clearly visible.

The range of the maximum absolute differences, $0.001^\circ$-$0.04^\circ$, shown in the bottom rows of the subplots in Figure~\ref{fig:angles}(a)-(d) is smaller than the range, $0^\circ$-$1^\circ$, of angular difference in spacecraft position vectors in the same coordinate system from two different data providers found in section~\ref{sect:comparisons_ephemeris}. The position vectors can differ for two reasons: differences in the source of the position data and differences in the data providers' software libraries, if a transformation is needed. To determine the magnitude of each, one must analyze the source code used to generate the positions, which is not readily available.

In summary, in this section, we have shown that differences in library implementations of reference frames range from $0.001^\circ$ to $0.04^\circ$ for the chosen set of coordinate frames and comparison axis.
This range of differences is comparable to the range of angles identified in section~\ref{sect:comparisons_ephemeris}, which includes the average angular change in Earth's dipole $Z$-axis over the past 45 years, the precession of Earth's rotation axis, changes in nutation of Earth's rotation axis over one year, and the precession of the ecliptic. The implication is that intercomparison of measurements transformed with different libraries may require an analysis of the differences in the transform implementations if the differences in the measurements are not significantly larger than these differences. This type of analysis is challenging because it requires access to the transform software, which may not be maintained or available, and because calculations are needed to quantify differences due to transform implementation choices.

\clearpage

\clearpage

\section{How transform information is created for spacecraft missions}
\label{sect:missions}

Archival data for space physics spacecraft missions typically include vector measurements in multiple reference frames. For NASA missions, archival data products are delivered to the Space Physics Data Facility \cite{SPDF}. Modern missions typically take one of two approaches for transforms: (a) A mission operations team develops SPICE kernels for select space physics reference frames, and SPICE software is used for the transforms; or (b) A non-SPICE software library is used for the transforms, for example, ROCOTLIB and SpacePy.

There is significant overlap in effort across missions; typically, a scientist familiar with space physics reference systems but not with their implementation options will use available software or SPICE kernels and develop tests. As noted in the conclusions, this effort to develop and verify transform calculations would not be necessary if either a database of transform matrices or versioned SPICE kernels were available that were accepted as a community standard.

For the SSCWeb data service \cite{SSCWeb}, which provides spacecraft positions, three approaches have been taken in order of priority (see \citeA{SSCWebProvenance} for a table indicating which approach was taken and provenance):

\begin{enumerate}
    \parskip 0.1in 

    \item Mission--developed SPICE kernels for positions that are obtained from the mission team periodically and archived. The \texttt{GEI} position--related SPICE kernel information is used to generate the \texttt{J2000} spacecraft positions. Then, software developed as part of the International Solar-Terrestrial Physics program in the 1990s (ISTP; \citeA{ISTP2025}) is used to compute positions in different coordinate frames, including \texttt{GEI}, \texttt{GEO}, \texttt{GM}, \texttt{GSE}, \texttt{GSM}, and \texttt{SM}. For ISTP--era missions, i.e., ACE, Polar, Wind, and Geotail, information needed for the reference frames, such as the position of the Sun, was calculated by GSFC's Flight Dynamics Facility (FDF) using proprietary software.

    \item A mission--computed \texttt{GEI} position--related variable in archival data files delivered to SPDF is then used to compute orbit and attitude data in different reference frames; only data in the \texttt{GEI}--related frame is used. This is a possible source of the differences shown in Figure~\ref{fig:geotail} if the mission computed positions in the archival data products by transforming \texttt{GEI} with software other than that used by SSCWeb. Differences can also occur if the mission uses a different definition of \texttt{GEI} than SSCWeb.

    \item If SPICE kernels or other mission--computed  \texttt{GEI} position-related values are not available, TLE (Two-Line Element) files are obtained from NORAD (North American Aerospace Defense Command), and a translation of the C NORAD library to Pascal \cite{NORADSGP4c} is used to compute positions in \texttt{TEME} (True Equator, Mean Equinox). Often, TLEs are used at the start of the mission, and a switch is made to approaches 1. or 2. as information becomes available. This switching has implications for reproducibility that are discussed in section~\ref{sect:ephemeris-version} because previous versions of positions are not retained.

\end{enumerate}

\section{Summary and Recommendations}
\label{sect:conclusions}

In this section, we provide recommendations to address the issues identified in this article. Updates on activities related to these recommendations will be available on the mailing list \texttt{https://groups.io/g/hdrl-stct}.

All of these recommendations should be addressed by an international standards organization. At this point, we do not recommend any single organization but note that relevant organizations include:

\begin{itemize}

    \item The Committee on Space Research \cite{COSPAR}, which issues resolutions on data and metadata standards;

    \item The International Association of Geomagnetism and Aeronomy \cite{IAGA2026}; the IAGA V-MOD Working Group \cite{IAGA_V-MOD2026} is responsible for the IGRF model specification;

    \item The International Earth Rotation and Reference Systems Service \cite{IERS}, which provides data for Earth orientation, the International Celestial Reference System/Frame (ICRS/ICRF), and the International Terrestrial Reference System/Frame (ITRS/ITRF) and also maintains the IERS Conventions, which include models, constants, and standards;

    \item The International Heliophysics Data Environment Alliance \cite{IHDEA}, which encourages the use of common standards and services;

    \item The International Virtual Observatories Alliance \cite{IVOA}, which has the Astronomical Coordinates and Coordinate Systems metadata model and a reference frame vocabulary; and

    \item Space Physics Search and Extract (SPASE;  \citeA{Roberts2018}), which has a list of definitions of coordinate frame acronyms.
\end{itemize}

\subsection{Standards for Reference System and Frame Definitions}
\label{sect:standard-for}

As defined in section~\ref {sect:terminology}, an {\it ideal reference system} is a general description of a reference frame that lacks key details necessary for its implementation. Examples were provided in which the \texttt{GEI} acronym had different expansions, and different acronyms were identified as equivalent.
  
The next level of specificity relative to an ideal reference system is a {\it reference system}, which defines constants, models with free parameters, and algorithms for transforming between observable quantities and reference data. 

Finally, a {\it reference frame} is a realization of a reference system based on measurements used to determine the free parameters in its models. 

These terms are used in the astrophysics and Earth sciences communities, and we recommend that the space physics community adopt them in place of the ambiguous term ``coordinate system'' (see section~\ref{sect:definitions}).

A community-developed standard for acronyms and definitions should be created. The standard should provide abbreviation/definition pairs for commonly used space physics ideal reference frames, so that data providers and paper authors can use them and reduce ambiguity. Associated with each ideal reference frame definition, there should be at least one reference system abbreviation/definition pair. If multiple reference systems are in common use, their definitions and referencing conventions should be specified. For each reference system, at least one reference frame should be defined.

The reference system definitions will involve key quantities, such as the position of the Sun and Earth's dipole orientation. As a result, these key quantities will need to be precisely defined along with the model and methodology for computing them.

Although this article did not provide details on the relevance of time standards to reference frame transforms, this topic is an important part of the definition of a reference frame. Several space physics reference frames depend on time-related quantities. Both \citeA{Hapgood1992} and \citeA{Franz2002} discuss the relevance of time scales to reference-frame calculations and approximations that may be used (for example, using UTC in place of UT1). Even if using an approximation is justified for the precision needed in most calculations, we must account for the fact that, when defining a reference frame, all quantities used in the calculation must be explicitly specified. \add[r2]{New standards should also address the planned discontinuation of the leap second \mbox{\cite{Gibney2022,Palmer2023}}, which will affect approximations such as the use of UTC in place of UT1.}

There exist standards that reference frame definitions can be built upon. For example, in the past 50 years, the astronomy community, through International Astronomical Union (IAU) resolutions, has adopted standards related to quantities used in space physics reference frame transforms in 1982~\cite{Aoki1982}, 2000~\cite{Capitaine2003}, and 2006~\cite{Capitaine2005}. The IGRF model \cite{IAGA_V-MOD2026} is a standard for specifying Earth's dipole orientation, which is required for defining geomagnetic reference frames.

\subsection{Reference Frame Standard Dataset}
\label{sect:reference-dataset}

For a given transform between reference frames, a transformation matrix at a specific instant in time (or, equivalently, rotation angles or quaternions) relative to a base reference frame is required. We also note that implementing transforms requires technical expertise in software and maintenance, including updating free parameters in reference-frame models, verifying calculations, and documenting changes. Although providing a software version enables reproducibility in principle, in practice, not all users will be familiar with or want to use the cited software, and in the long term, the software may not be maintained or may be difficult to install and use. 

Based on this, we recommend that a standards body develop a standard dataset with a long-term plan to keep it up-to-date. Each record in the dataset should consist of a timestamp and a transform matrix that relates a coordinate frame to a reference coordinate frame. The records should be made available in a standard scientific file format and from a web service API (Application Programming Interface). A parsimonious alternative to providing 9 transform matrix values is to define only one or two angles (and associated rotation axes) from which the matrix can be computed.

With such a dataset, if a scientist wants to allow for reproducibility, they could state, for example, ``the vector measurements in \texttt{ABC} were transformed into reference frame \texttt{XYZ}," \texttt{ABC} to \texttt{REF} and \texttt{REF} to \texttt{XYZ} using transformation matrices in the standard dataset and cite both the standard that defines the acronyms and the standard dataset.

Many software libraries do spot checks of transforms. For example, SunPy's unit tests \cite{SunPy} involve a comparison with a table from \citeA{Franz2002} at several time instants; \texttt{cxform} \cite{cxform} tests its implementation by comparing the positions in \texttt{GSE} of one year of data from three spacecraft from \citeA{SSCWeb}; a comparison is also made with a table in \citeA{Franz2002}. With the proposed reference database, a software library developer could make a more comprehensive comparison, allowing users to estimate uncertainties associated with coordinate transforms directly. An additional advantage of this dataset is that software developers can use it for more comprehensive testing of their implementations.

We also suggest that this dataset contains common time representations. Each record should be a UTC timestamp along with values for \texttt{UT1}, \texttt{TAI}, \texttt{TT}, and \texttt{ET} (\citeA{McCarthy2011}; see section~\ref{sect:glossary} for definitions) as well as astronomical time representations that are used in reference frame transforms, such as GMST.

The advantage of the standard dataset over software libraries is that the latter require continual upgrades, and not all users perform their analysis in the language in which the library is implemented. Although software that automatically updates IGRF coefficients and leap-second information may work now, it is not guaranteed to be maintained and usable indefinitely. In addition, transform results may change as a given software package evolves. For example, consider a package that utilizes non-definitive IGRF model coefficients for \texttt{MAG} and computes a transformation. Five years later, the result will not be the same if the software then uses definitive IGRF coefficients. Although, in principle, the package developers can maintain reproducibility, doing so requires significant effort. The proposed standard dataset could address this by defining two \texttt{MAG} reference frames. One uses only definitive IGRF coefficients. The other only uses non-definitive. We also recommend that each release of the IGRF coefficients be given its own DOI and archived in a generalized repository. (At present, it is only possible to cite descriptions of the model, e.g., \citeA{Alken2021}, but not a specific release.)

Having a dataset of reference frame transform matrices will address another issue. In the development of section~\ref{sect:comparisons_software}, we encountered significant errors in four of the libraries, despite these libraries passing their tests. (These errors were reported, and a version with the corrections was used in section~\ref{sect:comparisons_software}). If these libraries had a more comprehensive set of test matrices, these errors would have been noticed.

\subsection{Database of SPICE transform kernels}

Ideally, all mission-specific SPICE kernels would be available from a a collaboratively managed master repository with copies available at multiple locations, kernels related to standard space physics reference frames would be shared across missions, and these shared kernels would either be used to generate the standard dataset described in section~\ref{sect:reference-dataset} or an evaluation of the differences between transforms using these SPICE kernels and the standard dataset would be provided as documentation.

We also recommend using modern version control systems for the SPICE kernels. At present, kernels can be found on various websites, and kernel versions are sometimes indicated in the file name, but prior versions are not always provided, thereby inhibiting reproducibility. Storing kernels in a source code repository, which provides versioning and simplifies change tracking, will improve this. Additionally, we recommend assigning unique identifiers, such as DOIs, to kernels to facilitate their citation.

\subsection{On the use of $R_E$}
\label{sect:R_E}

There is no consensus on the definition of $R_E$. For example, \citeA{SSCWeb} uses 6,378.16~km, \citeA{Franz2002} uses 6,378.14~km. References to these two values and others may be found in \citeA{NASA1972}. The Automated Multi-Dataset Analysis (AMDA) system \cite{Genot2021} and SpacePy \cite{SpacePy} use 6,378.137~km. The value of 6,378.137~km is the semi-major axis length of the reference ellipsoid in the 1984 World Geodetic System (WGS 84) and the 1980 Geodetic Reference System (GRS80) standards. The IRBEM software library \cite{IRBEM2022} uses 6,371.2~km, which is the mean Earth radius used in the IGRF \cite{IAGA_V-MOD2026} model.

In the analysis of spacecraft positions described in section~\ref{sect:comparisons_ephemeris}, we encountered particular challenges when positions were given in Earth radii, $R_E$. In one case, the value of $R_E$ used was available, but difficult to find. In another case, the value of $6400$~km was claimed in metadata, but the actual value used was $6378.137$~km.

As a result, we recommend that data are never provided in terms of $R_E$, or, if this is not an option, data are also provided in the same dataset with units of m or km. In principle, metadata can provide a definitive value for $R_E$, but in practice, metadata is sometimes wrong or becomes detached from the data.

\subsection{Versioning and Documenting Positions}
\label{sect:ephemeris-version}

In this work, we have been primarily interested in the problem of transforming a vector from one reference frame to another. In the case of positions provided in a given reference frame, there is an uncertainty due to what is meant by that reference frame, which is addressed by the recommendation in section~\ref{sect:standard-for}. 

There is an additional source of uncertainty --- the source of the positions. Missions have used direct Doppler and GNSS measurements, \add[r2]{active ranging,} commercial software such as STK (Systems Tool Kit), and data from facilities such as the GSFC Flight Dynamics Facility (FDF), the Deep Space Network (DSN), NAIF, and NORAD to produce the spacecraft positions for the mission.

Accurate sources of spacecraft position data are also important -- even if two researchers compute the location of a spacecraft in a given reference using the same transformation algorithm, their results can differ significantly if the source of the position data differs. In section~\ref{sect:comparisons_ephemeris}, we gave an example in which differences in reported spacecraft positions between two data providers were likely due to one using mission-provided positions and the other using two-line elements that did not capture a spacecraft maneuver. \citeA{Hapgood2026} found that two-line elements from SpaceTrack map to the wrong Cluster spacecraft in about 10\% of the cases, and the two-line element best-matched perigee time with the ESA Joint Operations Science Center value differed by up to 60~sec in the time interval 2000-09-05 to 2024-08-27. In general, \change[r2]{the mission has access}{modern mission teams have access} to positions derived from the most accurate tracking data, including direct Doppler and on-board GNSS measurements, which are more accurate than two-line element-derived positions and may include spacecraft maneuvers that are not resolved by two-line elements.

To address this, we suggest that data providers indicate how they obtained their spacecraft positions. We also suggest that if two data providers provide positions for the same spacecraft, they document how and why their results may differ, if applicable. We also recommend that the metadata indicate the uncertainty in the locations. Often, data providers will modify the position values when more accurate calculations become available. This makes reproducibility difficult. Ideally, for reproducibility, old versions of positions would be made available. Often, this is impractical, so an alternative is to include version information in the metadata, including a short description of the changes from the last version to the new one. The metadata for previous versions should be made available so users can understand the changes from earlier versions to the current one. Finally, the version number and its identifier should be included in the metadata.

\subsection{Documentation}

If the above recommendations are followed, the necessary documentation for reproducibility related to coordinate transforms will be simplified. For example, if an author or software package used the standard dataset described in section~\ref{sect:reference-dataset} for a transform, only the standard dataset is needed for reproducibility. Similarly, if one or more SPICE kernels are used, each with a unique DOI and stored in a permanent generalized repository, only the DOIs are needed to retrieve the kernels. In this case, these DOIs and the version of the SPICE software library used are sufficient for reproducibility.

We recommend the following for general software and data documentation.

\begin{itemize}
  \item Software package developers provide documentation on implementation choices (e.g., model used for obliquity and its parameters). If data from an external resource is used, for example, if the IGRF coefficients are dynamically downloaded, the user should have an option to easily view the data used, so that, if necessary, the values used can be documented in a publication.
  Software should have a DOI for each major version, so the version used can be referenced rather than just a top-level DOI. Additional software citation recommendations are given in \citeA{Jackson2012}, \citeA{Smith2016}, and \citeA{Stall2023}.

  \item Data providers ensure that metadata associated with positions and variables that are transformed include details about the software version used for the transforms or the data used for the transform, such as SPICE kernels. The metadata description should be validated by having a third party assess whether it contains sufficient information to re-implement the transform.

  \item When data are provided that have been transformed to a reference frame that depends on a magnetic field model (such as the IGRF model), data are also provided in a frame such as GEI/J2000 or ECI/J2000. The motivation is the variability in the implementation of the field model. 
  
  \item Paper authors include the version of software, a DOI for that version if available, and the options passed to the transform function. If SPICE kernels are used, they should be permanently publicly accessible and citable with a version-specific identifier. 

\end{itemize}

\appendix
\section{Definitions}
\label{sect:glossary}

\begin{itemize}
\item ACE -- Advanced Composition Explorer (spacecraft mission)
\item CDAWeb -- Coordinated Data Analysis Web (NASA data archive)
\item \texttt{CD} -- Centered Dipole (reference frame). Same as GM and MAG if they use a model dipole with an origin of Earth's center of mass.
\item CDF -- Common Data Format (file format for scientific data)
\item DSN -- Deep Space Network (NASA communications network)
\item ECI -- Earth-Centered Inertial (reference frame)
\item  \texttt{EME} -- Earth Mean Equator (reference frame)
\item  ET -- Ephemeris Time (time standard)
\item  \texttt{GEI} -- Geocentric Equatorial Inertial (reference system)
\item  \texttt{GEO} -- Geographic (reference system)
\item  \texttt{GM} -- Geomagnetic (reference system, often called MAG)
\item  \texttt{GNSS} -- Global Navigation Satellite System
\item  \texttt{GSE} -- Geocentric Solar Ecliptic (reference system)
\item  \texttt{GSM} -- Geocentric Solar Magnetospheric (reference system)
\item  \texttt{GCI} -- Geocentric Celestial Inertial (reference system)
\item  \texttt{GCRS} -- Geocentric Celestial Reference System (reference system)
\item GSFC -- Goddard Space Flight Center (NASA center)
\item IAGA -- International Association of Geomagnetism and Aeronomy
\item IGRF -- International Geomagnetic Reference Field (geomagnetic model); maintained by the IAGA (IAGA, 2026) V-MOD (IAGA/V-MOD, 2026) Working Group.
\item ISTP -- International Solar-Terrestrial Physics (NASA program)
\item IRBEM -- International Radiation Belt Environment Modeling (software library)
\item ICRF -- International Celestial Reference Frame (reference frame)
\item ICRS -- International Celestial Reference System (reference system)
\item  \texttt{J2000} -- Reference system with reference epoch of 2000 January 1 noon TT.
\item J2000.0 -- Reference epoch associated with  \texttt{J2000}
\item  \texttt{J2K} -- Abbreviation of  \texttt{J2000}
\item JPL -- Jet Propulsion Laboratory
\item  \texttt{MAG} -- Geomagnetic (reference system). Also referred to as Centered Dipole (\texttt{CD}) and \texttt{GM}.
\item MMS -- Magnetospheric Multiscale Misson 
\item NAIF -- Navigation and Ancillary Information Facility 
\item MOD -- Mean of Date (reference system modifier, precession only)
\item NORAD -- North American Aerospace Defense Command
\item  \texttt{SM} -- Solar Magnetic (reference system)
\item SPICE -- Spacecraft, Planet, Instrument, C-matrix, Events (NAIF toolkit)
\item SPEDAS -- Space Physics Environment Data Analysis Software (software library)
\item SSCWeb -- Satellite Situation Center Web (NASA data service)
\item STK -- Systems Tool Kit (commercial orbit analysis software)
\item TAI -- International Atomic Time (time standard)
\item TT -- Terrestrial Time (time standard)
\item TLE -- Two-Line Element set (satellite orbital elements)
\item  \texttt{TEME} -- True Equator, Mean Equinox (reference system)
\item UT1 -- Universal Time 1 (time standard)
\item UTC -- Coordinated Universal Time (time standard)
\end{itemize}

\section*{Open Data}

The software for the associated calculations and plots is available in \citeA{Weigel2026}. The calculations were made using the \texttt{hxform} package \cite{Weigel2025}, which is an interface to the software packages and libraries used in section~\ref {sect:comparisons_software}.

\section*{Acknowledgments}

This work was supported by NASA Grant 80NSSC24K0199. We thank Fernando Carcaboso Morales, Lan Jian, Rita Johnson, Leonard Garcia, Tami Kovalick, Boris Semenov, Jon Niehof, Steve Morely, Edmund Henley, and Nathaniel Bachman for feedback on the manuscript or additional background information, and Jim Lewis for addressing questions about PySPEDAS. We also thank the reviewers, who include an anonymous reviewer and M. Hapgood, who provided extensive feedback and suggestions that have improved this manuscript.

\section*{Conflict of Interest declaration}

\noindent
The authors declare there are no conflicts of interest for this manuscript.

{\raggedright \bibliography{main}}

\end{document}